\documentclass[usenatbib]{mn2e}
\usepackage{graphicx}
\usepackage{epsfig}

\def\gtsima{$\; \buildrel > \over \sim \;$}
\def\ltsima{$\; \buildrel < \over \sim \;$}
\def\gtrsim{\lower.5ex\hbox{\gtsima}}
\def\lesssim{\lower.5ex\hbox{\ltsima}}

\title[Perturbations on the GC stellar disc]
{Perturbations induced by a molecular cloud on the young stellar disc in the Galactic Centre}

\author[Mapelli et al.]  
  {Michela Mapelli$^{1}$\thanks{E-mail: michela.mapelli@oapd.inaf.it}, Alessia Gualandris$^{2}$ and Tristen Hayfield$^{3}$\\
  $^{1}$INAF-Osservatorio Astronomico di Padova, Vicolo dell'Osservatorio 5, I--35122, Padova, Italy\\
  $^{2}$Department of Physics and Astronomy, University of Leicester, Leicester, LE1 7RH, United Kingdom\\
  $^{3}$Max Planck Institute for Astronomy, K\"onigstuhl 17, D--69117 Heidelberg, Germany\\
   }

\begin{document}

\date{}

\maketitle

\begin{abstract}
The Galactic centre (GC) is a crowded environment: observations have revealed the presence of (molecular, atomic and ionized) gas, of a cusp of late-type stars,  and of $\sim{}100$ early-type stars, about half of which lying in one or possibly two discs. In this paper, we study the perturbations exerted on a thin stellar disc (with outer radius $\sim{}0.4$ pc)  by a molecular cloud that falls towards the GC and is disrupted by the supermassive black hole (SMBH). The initial conditions for the stellar disc were drawn from the results of previous simulations of molecular cloud infall and disruption in the SMBH potential. We find that most of the gas from the disrupted molecular cloud settles into a dense and irregular disc surrounding the SMBH. 
 If the gas disc and the stellar disc are slightly misaligned ($\sim{}5-20^\circ{}$), the precession of the stellar orbits induced by the gas disc significantly increases the inclinations of the stellar orbits  (by a factor of $\sim{}3-5$ in 1.5 Myr)  with respect to the normal vector to the disc. Furthermore, the distribution of orbit inclinations becomes significantly broader. 
These results might be the clue to explain the broad distribution of observed inclinations of the early-type stars with respect to the normal vector of the main disc. We discuss the implications for the possibility that fresh gas was accreted by the GC after the formation of the disc(s) of early-type stars.
\end{abstract}


\section{Introduction}
The compact radio source Sgr~A$^\ast{}$, located at the very centre of our Galaxy, coincides with a high concentration of mass ($\approx{}4\times{}10^6$ M$_\odot$), 
 almost certainly a supermassive black hole (SMBH,  \citealt{genzel03}; \citealt{schodel03}; \citealt{ghez03}, 2005).
More than a hundred young massive stars have been observed in the vicinity of Sgr~A$^\ast{}$ (\citealt{krabbe91}; \citealt{morris93}; \citealt{krabbe95}; \citealt{genzel03}).  Many of them are O-type and Wolf-Rayet (WR) stars. About half of the early-type stars lie in a thin disc with radius 0.04 pc$\,{}\lesssim{}r\lesssim{}0.5$ pc and average eccentricity $e\sim{}0.3-0.4$ (\citealt{bartko09}; \citealt{lu09}; \citealt{yelda12}; \citealt{do13}; \citealt{lu13}). This disc is called clock-wise (CW) disc, because it shows CW motion when projected on the plane of the sky (\citealt{genzel03}; \citealt{paumard06}).  The CW disc is likely warped, as the orientation of its normal vector changes by several degrees ($\sim{}60^\circ{}$, according to \citealt{bartko09}) from its inner to its outer edge. A fraction of the remaining early-type stars show counterclockwise motion, which may indicate the presence of a second dissolving disc (\citealt{lu06}, 2009; \citealt{bartko09}). The age of  the observed early-type stars is  $t_{\rm age}\approx{}2.5-6$ Myr (\citealt{lu13}; see \citealt{paumard06} for a previous estimate $t_{\rm age}=6\pm{}2$ Myr). Furthermore, \citet{yusef-zadeh13} found indications of gas outflows, suggesting recent star formation (10$^{4-5}$ yr) within 0.6 pc of SgrA$^\ast$.  The $\sim{}20$ stars closest to SgrA$^\ast{}$ ($\lesssim{}0.04$ pc $\sim{}1$ arcsec) are B stars, with an age 20-100 Myr. These, named the S-stars, have very eccentric and randomly oriented orbits (\citealt{schodel03}; \citealt{ghez03}, 2005; \citealt{eisenhauer05}; \citealt{gillessen09}). The ensemble of the (both young and old) stars in the central few parsecs is often referred to as the nuclear star cluster (NSC) of the Milky Way (MW). 

The Galactic centre (GC) is a very crowded environment not only for the stellar population, but also for the gas. A molecular torus, the circumnuclear ring (CNR, \citealt{genzel85}; \citealt{gusten87}; \citealt{yusef-zadeh04}; \citealt{christopher05}; \citealt{oka11}), is located at $\sim{}2$ pc from SgrA$^\ast$, and is on the verge of forming stars (\citealt{yusef-zadeh08}). Furthermore, the innermost 3 pc of the GC are extremely rich in ionized gas (e.g., \citealt{zhao09}, and references therein). Finally, two molecular clouds (the M--0.02--0.07 and the M--0.13--0.08 cloud, \citealt{solomon72}; \citealt{novak00}) lie within 20 pc of the GC.

The origin of the early-type stars is puzzling, as the strong tidal field exerted by the SMBH is expected to disrupt any molecular cloud before it reaches the distance of the CW disc. On the other hand, 
a disrupted molecular cloud might spiral towards the SMBH and form a gas disc around it, sufficiently dense to fragment into stars 
(\citealt{levin03}; \citealt{goodman03}; \citealt{goodman04}; \citealt{milosavljevic04}; \citealt{nayakshincuadra05}; \citealt{rice05}; \citealt{alexander08}; \citealt{collin08}; \citealt{bonnell08}; \citealt{mapelli08}; \citealt{wardle08}; \citealt{hobbs09}; \citealt{alig11}; \citealt{jiang11}; \citealt{namekata11}; \citealt{mapelli12}, hereafter M12; \citealt{lucas13}; \citealt{alig13}). %
In particular, M12 simulated the infall of a molecular cloud towards the GC. In the M12 simulations, the molecular cloud is disrupted by the SMBH tidal shear, and settles into a dense gas disc, which fragments into self-gravitating clumps.
%
These clumps, or proto-stars, lie in a disc at a distance of 0.1--0.4 pc with moderately eccentric orbits ($e\sim0.2-0.4$). The properties of the stellar disc reproduce quite well the observations of the CW disc, but cannot explain the counterclockwise stars as well as the S-stars. On the other hand, M12 integrates the evolution of the cloud only for $\sim{}0.5$ Myr, while the early-type stars are  $\sim{}2.5-6$ Myr old (\citealt{lu13}).

The GC is such a crowded environment that many external forces might have influenced the orbital evolution of the CW disc after its formation. First, the stellar cusp of late-type stars induces precession on the orbits of a stellar disc (e.g. \citealt{lockmann09a}). This is expected to significantly affect the eccentricity distribution (\citealt{madigan09}; \citealt{gualandris12}, hereafter G12). Furthermore, a massive perturber, such as an intermediate-mass black hole (\citealt{gualandris09}; \citealt{gualandris10}; \citealt{perets10}), the CNR (\citealt{haas11a}; \citealt{haas11b}), a second stellar disc (\citealt{lockmann08}; \citealt{lockmann09b}; \citealt{lockmann09a}), or a second molecular cloud infalling towards SgrA$^\ast$ is expected to induce a precession on the CW disc, and it may also excite the Kozai resonance (\citealt{kozai62}; \citealt{lidov62}). In this paper, we simulate the infall of a second molecular cloud toward the SMBH, and we study the influence that the second cloud has on the orbits of a pre-existing stellar disc. As initial conditions for the  pre-existing stellar disc, we adopt the outcomes of run~E in M12.


\section{N-body simulations}
For our simulations, we used the  N-body/smoothed particle hydrodynamics (SPH) code \textsc{gasoline} (\citealt{wadsley04}), upgraded with the \citet{read10} optimized SPH (OSPH) modifications, to address the SPH limitations outlined, most recently, by \citet{agertz07}. 
\begin{figure}
  \center{
    \epsfig{figure=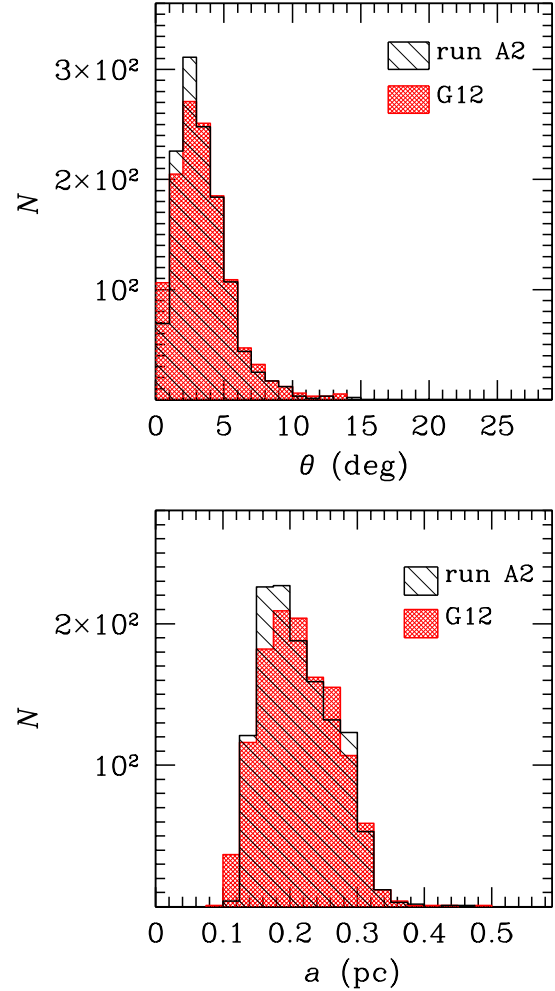,width=8.0cm} 
  \caption{  \label{fig:fig1}
  Top (bottom): distribution of inclinations (semi-major axes) at $t=1.5$ Myr. The inclinations are measured by the angle $\theta{}$, between the normal vector to the orbit of a single star and the total angular momentum vector of the stellar disc (see Appendix~A). Cross-hatched red histogram: simulation from G12. Hatched black histogram: run~A2.}}
\end{figure}

Table~1 shows a summary of the runs that will be presented in this paper. In all the runs, we simulate a disc of 1252 star particles orbiting the SMBH. The SMBH is represented by a sink particle, with initial mass $M_{\rm SMBH}=3.5\times{}10^6$ M$_{\odot{}}$ (\citealt{ghez03}), sink radius $r_{\rm acc}=5\times{}10^{-3}$ pc and softening radius $\epsilon{}=1\times{}10^{-3}$ pc. 

The disc of star particles is obtained from run~E of M12, with the same procedure as described in G12. In particular, we assume that each self-bound clump formed in run~E of M12 becomes a star, without accreting any further gas particles after $t=4.8\times{}10^5$ yr\footnote{This assumption is quite conservative for the mass of stars, as it is realistic to expect further gas accretion in run~E of M12.}. Thus, we replace each gas particle with a single star particle having mass equal to the total mass of the gas clump, position and velocity corresponding to those of the centre-of-mass of the clump. The resulting simulated mass of the disc is $4.1\times{}10^3$ M$_{\odot{}}$ (see M12). The best-fitting initial mass function (IMF) is given by a single power law with index $\alpha{}=1.50\pm{}0.06$ (see M12 for details), consistent with Lu et al. (2013, $\alpha{}=1.7\pm{}0.2$). Considering that M12 simulations cannot resolve stellar masses $<1.3$ M$_{\odot{}}$, this corresponds to an expected total mass of the disc $\approx{}4.5\times{}10^3$ M$_{\odot{}}$ (assuming $\alpha{}=1.50$). This value is consistent with most observations (e.g. \citealt{paumard06}; \citealt{bartko09}), but is a factor of $\sim{}3$ lower than the most recent results (\citealt{lu13}).  Finally, the stellar disc has average semi-major axis $\langle{}a\rangle{}=0.21\pm{}0.04$ pc, average inclination\footnote{For the definition of the coordinate system, see Appendix A.} $\langle{}{\theta{}}\rangle{}=2.4\pm{}1.5^\circ{}$, and average eccentricity $\langle{}e\rangle{}=0.29\pm{}0.04$.

 The high density and the clumpiness of the gas disc at $t=4.8\times{}10^5$ yr in run~E of M12 
make prohibitive to continue the simulation as it is. Thus, in our initial conditions, we instantaneously removed the gas left in the first disc, to prevent our simulations from stalling before the infall of the second cloud. For the issues related to this assumption, see the discussion in Section~\ref{issue}. 

In runs A2 and B2, we also add a rigid potential, to account for the stellar cusp surrounding Sgr~A$^\ast{}$. The overall density profile of the stellar cusp goes as $\rho{}(r)=2.8\times{}10^6\,{}{\rm M}_{\odot{}}\,{}{\rm pc}^{-3}\,{}(r/\textrm{0.22 pc})^{-\gamma}$, where $\gamma=1.2$ (1.75) for $r<0.22$ pc ($r>0.22$ pc), consistent with the values reported in \citet{schodel07}. Runs~A2 and B2 will be referred to as runs `with cusp', while runs A1 and B1 will be dubbed runs `without cusp'.

\begin{table}
\begin{center}
\caption{Initial conditions.} \leavevmode
\begin{tabular}[!h]{llll}
\hline
Run
& Stellar Cusp$^{\rm a}$
& Perturber$^{\rm b}$
& $\alpha{}$ ($^\circ{}$)$^{\rm c}$\\
\hline
\noalign{\vspace{0.1cm}}
A1       & No  & No & -- \\
A2       & Yes & No & -- \\
\noalign{\vspace{0.1cm}}
B1       & No  & Yes & 0 \\
B2       & Yes & Yes & 0 \\
\hline
\end{tabular}
\end{center}
\footnotesize{$^{\rm a}$ The stellar cusp is the rigid potential corresponding to the density distribution in \citet{schodel07}. $^{\rm b}$ The perturber is a MC falling towards the GC.   $^{\rm c}$ $\alpha{}$ is the inclination between the initial orbit of the MC (when present in the simulation) and the plane of the stellar disc. $\alpha{}=0$ means that the orbit of the cloud lies in the plane of the stellar disc.} 
\end{table}


In runs B1 and B2, we simulate the infall of a second molecular cloud towards the GC, to study the perturbations induced by the cloud on the orbits of the stellar disc. In runs~A1 and A2, we do not include the second molecular cloud: the stellar disc evolves under the influence of the SMBH, and, in the case of run~A2, also under the influence of the rigid stellar cusp. Runs A1 and A2 were performed just for comparison with the other two runs.

The perturber cloud is simulated as in M12. In particular, it is a spherical cloud with a radius of 15 pc and a mass of $1.3\times{}10^5$ M$_{\odot{}}$. The cloud is seeded with supersonic turbulent velocities and marginally self-bound (see \citealt{hayfield11}). 
The centre-of-mass of the cloud is initially at 25 pc from the SMBH. The orbit of the cloud was chosen so that the impact parameter with respect to the SMBH is $10^{-2}$ pc and the initial velocity is one tenth of the escape velocity from the SMBH at the initial distance (i.e. the orbit is bound and highly eccentric). The orbit of the cloud lies in the plane of the stellar disc ($\alpha{}=0$, see Table~1).
As in run~E of M12, we include radiative cooling in all our simulations. The radiative cooling algorithm is the same as that described in \citet{boley09} and in \citet{boley10}. 
\citet{dalessio01} opacities are used, with a 1~$\mu{}$m maximum grain size. The irradiation temperature is $T_{\rm irr}=100$~K everywhere. The mass of the gas particles in each simulation is 0.4~M$_{\odot{}}$ and the softening length $10^{-3}$~pc. Given the low mass resolution, these simulations cannot be used to study gas fragmentation. In a forthcoming paper, we will show new high-resolution runs focused on this aspect, while in the current paper we are interested in the role of the cloud as perturber of the pre-existing stellar disc.


The code used for our simulations adopts a kick-drift-kick leapfrog integrator. This scheme is known to be rather inaccurate in the short-term integration of the orbits, while it preserves well energy and angular momentum on the long term evolution (e.g., \citealt{zemp07}; \citealt{dehnen11}). This may damp any secular changes in eccentricity (e.g., \citealt{zemp07}). 
On the other hand, the adopted code is the best compromise between describing the orbits accurately and following the thermodynamical and dynamical evolution of a living cloud (rather than assuming a rigid potential for the gas, as done in previous work).

We checked the limits of our numerical approach by comparing them with the results of G12, who used a Hermite integration scheme (implemented in the direct-summation N-body code $\phi{}$\textsc{GRAPE}, \citealt{harfst07}). In Fig.~\ref{fig:fig1}, the distribution of semi-major axes and inclinations obtained at $t=1.5$ Myr by G12 are compared with the same distributions derived at $t=1.5$ Myr for our run A2 (whose main ingredients are the same as in G12).  No evident differences appear between G12 and our run A2, by looking at Fig.~\ref{fig:fig1}. The Kolmogorov-Smirnov test indicates a 80 (30) per cent probability that the distribution of semi-major axes (inclinations) is the same in G12 and in run A2. Such probabilities are quite high, considering the intrinsic differences of the two runs. 
Other checks (e.g. reducing the time stepping, changing the number of neighbours, substituting the SMBH particle with a rigid potential, integrating some analytic orbits) demonstrated that our results are robust and fairly describe the effects of precession on the stellar orbits.  Furthermore, stochastic fluctuations (between different realizations of run~A2) change the orbital parameter distributions by less than $\sim{}10$ per cent (i.e. do not significantly affect the mean values of the orbital parameters).

The simulations presented in this paper ran on the Fermi IBM Blue Gene/Q at CINECA (through CINECA Award N. HP10CL51UF, 2012). Runs~B1 and B2 required approximately 100k CPU hours each.

\section{Results}

\subsection{The disruption of the cloud and the formation of the gas disc}\label{gasev}
\begin{figure}
  \center{
    \epsfig{figure=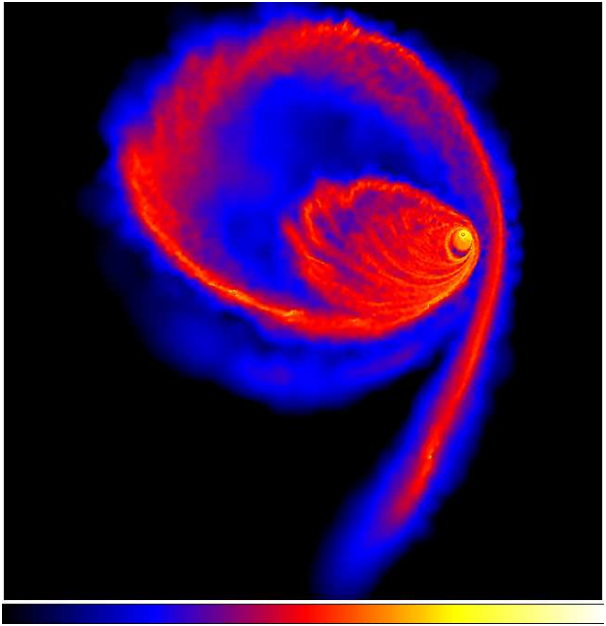,width=8.0cm} 
  }
  \caption{  \label{fig:fig2}
Projected density of gas in run~B1 at $t=1.5$ Myr. The gas disc is seen face-on. The box measures 35 pc per edge. The color map is logarithmic, ranging from $2\times{}10^{-4}$ to $2\times{}10^{8}$ M$_\odot$ pc$^{-3}$.}
\end{figure}

\begin{figure}
  \center{
    \epsfig{figure=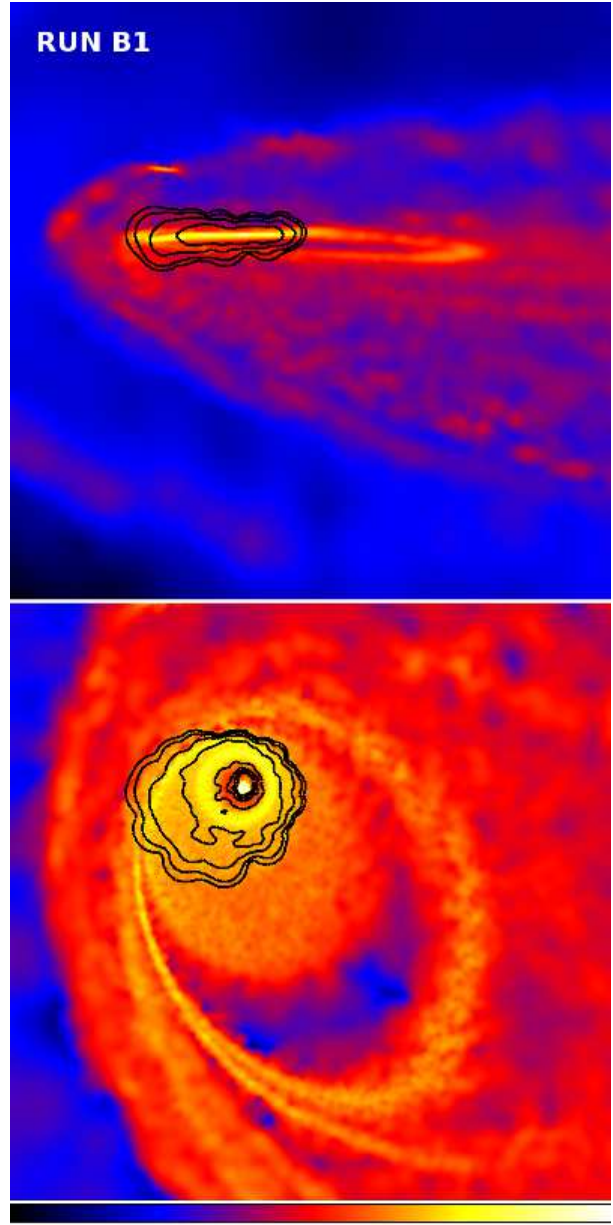,width=8.0cm} 
  }
  \caption{  \label{fig:fig3}
Projected density of gas in run~B1 at $t=1.5$ Myr.  The color map is logarithmic, ranging from $2\times{}10^{-4}$ to $2\times{}10^{8}$ M$_\odot$ pc$^{-3}$. The contours show the projected density of stars in the stellar disc. Top panel: the stellar disc is seen edge-on. Bottom-panel: the stellar disc is seen face-on. Each box measures 2.3 pc per edge.}
\end{figure}
 In our runs  B1 and B2, the second molecular cloud spirals towards the GC, and is quickly disrupted by the tidal force exerted by the SMBH. The disruption of the cloud starts at $t\approx{}10^5$ yr since the beginning of the simulations, i.e. $\approx{}6\times{}10^5$ yr after the infall of the first cloud (simulated in M12). At $t\approx{}3-5\times{}10^5$ yr  since the beginning of the simulation, a dense gaseous ring forms, surrounding the SMBH.  In the next Myr, the ring is fueled by fresh gas coming from the remnant of the disrupted molecular cloud (see Fig.~\ref{fig:fig2}). For this reason, the central gas disc looks more like a series of streamers than a coherent thin disc (see Fig.~\ref{fig:fig3}). We notice that the gravitational potential exerted on the cloud is slightly different in run~B1 and B2, since a spherical stellar cusp is included only in run~B2. Thus, the orbits of the streamers are slightly different in the two runs.


The initial inclination of the orbit of the cloud is nearly preserved during the evolution and the tidal disruption. In runs B1 and B2, the orbit of the second molecular cloud is in the plane of the stellar disc ($\alpha=0$), and  the gas disc and the pre-existing stellar disc are nearly coplanar (see Fig.~\ref{fig:fig3}). 

\begin{figure}
  \center{
    \epsfig{figure=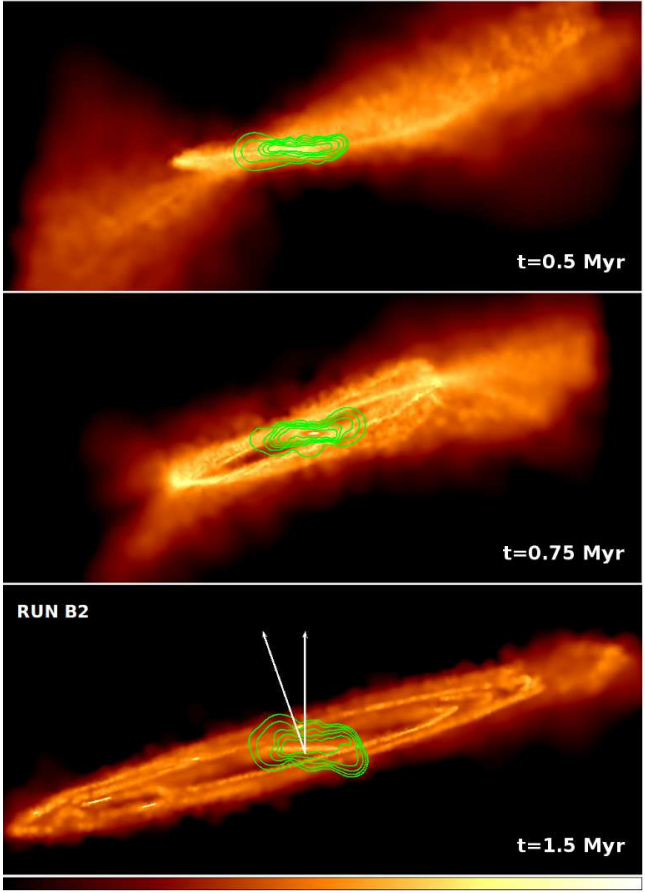,width=8.5cm} 
  }
  \caption{  \label{fig:fig4}
Projected density of gas in run~B2 at $t=0.5$, 0.75 and 1.5 Myr in the top, central and bottom panel, respectively.  The color map is logarithmic, ranging from $2\times{}10^{-2}$ to $2\times{}10^{10}$ M$_\odot$ pc$^{-3}$. The contours show the projected density of stars in the stellar disc. The box size is $4.0\times{}1.8$ pc. The projection was chosen so that the total angular momentum of the stellar disc is aligned to the vertical axis of the plot. The two white arrows in the bottom panel show the direction of the  total angular momentum of the stellar disc and the total angular momentum of the outer gas disc. The length of the arrows is arbitrary.}
\end{figure}

\begin{figure}
  \center{
    \epsfig{figure=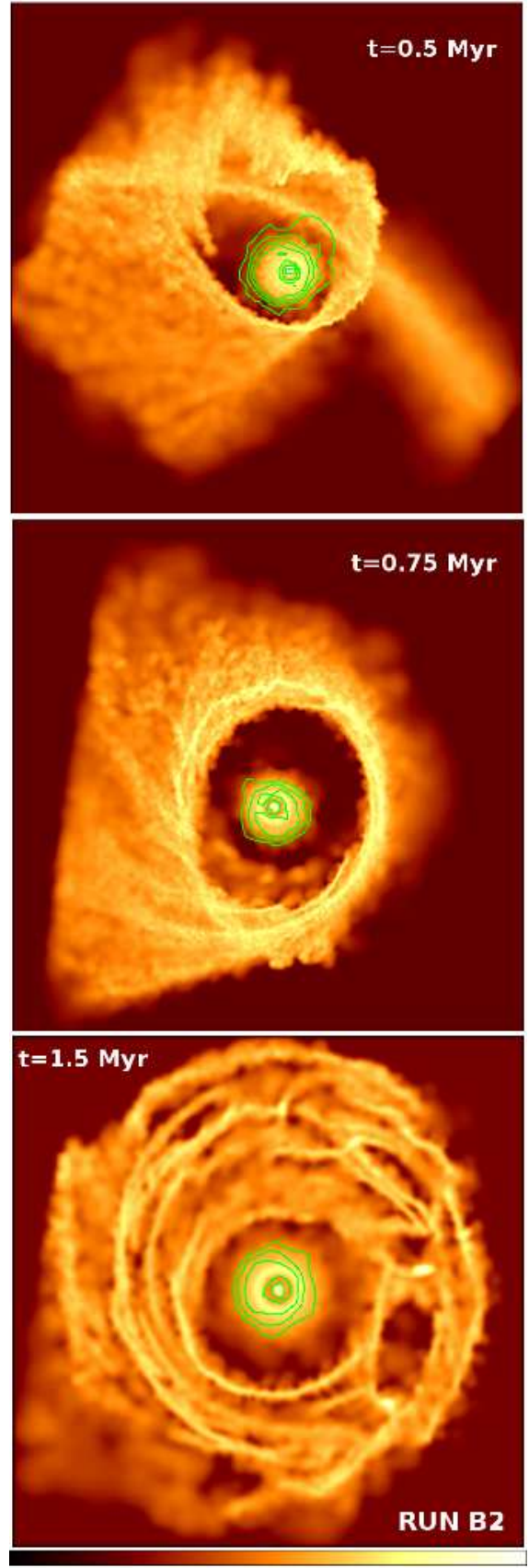,width=7.5cm} 
  }
  \caption{  \label{fig:fig5}
The same as Fig.~\ref{fig:fig4}, but rotated so that the stellar disc is seen face on. The box size is $4\times{}4$ pc. }
\end{figure}

On the other hand, the gas disc is not a rigid body, but is instead composed of many concentric annuli and streamers, all of which may have slightly different ($\sim{}5^\circ{}$) mutual inclinations (see Figs.~\ref{fig:fig3}, \ref{fig:fig4} and  \ref{fig:fig5}). Furthermore, the fact that the cloud is seeded with supersonic turbulence implies that clumps and filaments form in the molecular cloud before disruption. This determines local changes in the geometry of the infalling cloud, and is source of important stochastic fluctuations in the structure of the gas disc, between different simulations. In particular, we remind that the initial radius of the cloud is $\sim{}15$ pc, while the radius of the stellar disc is less than one pc: even small density fluctuations inside the cloud can significantly affect the geometry of the encounter between the stellar disc and the cloud.

In particular, Figs.~\ref{fig:fig4} and ~\ref{fig:fig5} show how the gas disc assembles between 0.5 and 1.5 Myr. We notice that the orbits of the streamers can change with time, as long as the disrupted molecular cloud re-fuels the inner disc with fresh gas. This makes the interplay between stellar disc and gas disc(s) very complicated.  It is apparent that while the stars define a thin and coherent disc, the gas is distributed in a series of streamers, with different angular momentum directions.

Finally, by the end of the simulation, when the cloud is entirely disrupted and the gas discs are completely  settled (i.e. the shocks and the cooling induced by the disruption are over and almost no new gas is accreted), the situation is even more complex. The bottom panels of Figs.~\ref{fig:fig4} and ~\ref{fig:fig5} show the inner 2 pc of run~B2 at $t=1.5$ Myr. It is apparent that there are at least two different gas rings: a inner one, with radius $\sim{}0.4$ pc (similar to the stellar disc), and a outer one, with radius $\sim{}1.5$ pc. While the inner gas disc is almost coplanar with the stellar disc, the angular momentum vector of the outer gas disc is inclined by $\sim{}15-20^\circ{}$ with respect to the angular momentum vector of the stellar disc. The outer gas ring may play the same role as the CNR observed in our Galaxy.


\begin{figure}
  \center{
    \epsfig{figure=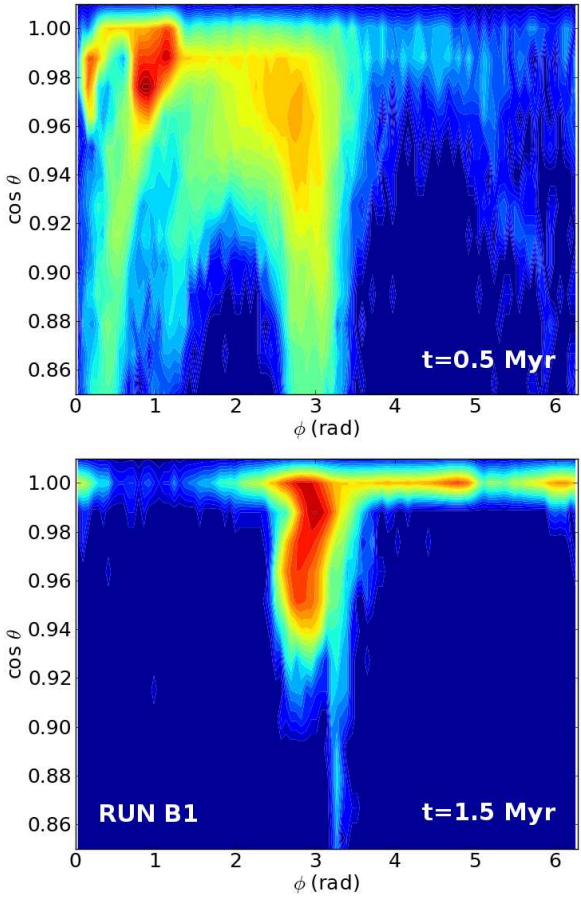,width=8.0cm} 
  }
  \caption{\label{fig:fig6}
Density of angular momentum vectors of gas in run B1, in the plane defined by  $\cos{\theta}$ and $\phi{}$ (see equation~\ref{eq:A2}). Top panel: $t=0.5$ Myr. Bottom panel: $t=1.5$ Myr. Only gas particles on bound orbits around the SMBH with semi-major axis $0.1\le{}a/{\rm pc}\le{}10$ are shown. The reference system is defined by the total angular momentum of the stellar disc at time $t$, as described in Appendix A. Note that the zero point of $\phi$ and $\theta{}$ is not exactly the same between the two panels. The colour map is logarithmic.}
\end{figure}
\begin{figure}
  \center{
    \epsfig{figure=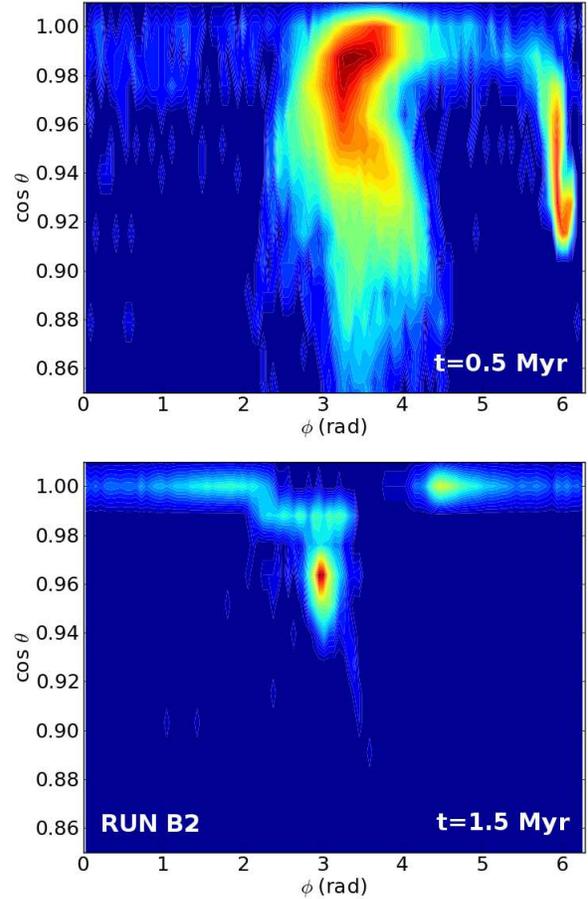,width=8.0cm} 
  }
  \caption{\label{fig:fig7}
The same as Fig.~\ref{fig:fig6}, but for run B2.}
\end{figure}

Figs.~\ref{fig:fig6} and \ref{fig:fig7} show the angular momentum distribution of gas particles moving on bound orbits around the SMBH (with semi-major axis $0.1\le{}a/{\rm pc}\le{}10$) in run~B1 and B2, respectively. The top and the bottom panel refer to time $t=0.5$ and 1.5 Myr, respectively. The coordinate system refers to the total angular momentum of the stellar disc at a given time $t$, as defined in Appendix~A. We remind that particles belonging to a razor thin disc aligned with the stellar disc have $\cos{\theta{}}=1$ and can assume any possible value of $\phi{}$. Instead, a razor thin disc that is misaligned with respect to the stellar disc defines an infinitely small circle in the $\cos{\theta{}},\,{}\phi{}$ plane. The larger the opening angle of the disc, the larger the circle it defines in the $\cos{\theta{}},\,{}\phi{}$ plane. The inclination between the stellar disc and a second misaligned disc is given by the difference in $\theta{}$. 

From Figs.~\ref{fig:fig6} and  \ref{fig:fig7} it is apparent that the gas discs are neither razor-thin nor regular structures. In particular, the top panels of Figs.~\ref{fig:fig6} and  \ref{fig:fig7}  show that the situation is rather complicated at $t=0.5$ Myr, i.e. during the first phase of the disc assembly (see also Figs.~\ref{fig:fig4} and \ref{fig:fig5}). At $t=1.5$ Myr the gas discs are more coherent, especially in the case of run~B2 (bottom panel of Fig.\ref{fig:fig7}), where the angular momentum vectors concentrate at $\cos{\theta{}}=0.964$ and $\phi{}=3$ rad. In both runs~B1 and B2, the total angular momentum direction of the main gas disc is similar to the one of the stellar disc, but is slightly offset. The offset at $t=1.5$ Myr is $\sim{}10^\circ$ and $\sim{}15^\circ$ in runs~B1 and B2, respectively. These issues are very important for the effects of the gas on the stellar orbits, as it will be discussed in Section~\ref{starev}.


\subsection{Timescales for precession}

\begin{figure}
  \center{
    \epsfig{figure=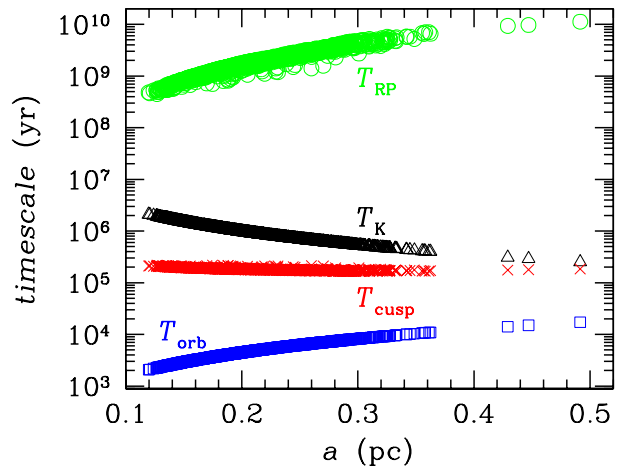,width=8.5cm} 
  }
  \caption{  \label{fig:fig8}
Comparison of the relevant timescales as a function of the semi-major axis for run B2. Open circles: $T_{\rm RP}$; open triangles: $T_{\rm K}$; crosses: $T_{\rm cusp}$; open squares: $T_{\rm orb}$.
}
\end{figure}

The two main sources of perturbation of the stellar orbital inclinations  in our simulations are: (i) the precession induced by the rigid stellar cusp; (ii) the precession induced by the disrupted molecular cloud. In this section, we give an estimate of the timescales associated with these precessions, for the quantities involved in our simulations.

(i) The  stellar cusp, i.e. a spherical potential, induces a precession of the orbits of the disc stars on a timescale (\citealt{ivanov05}; \citealt{lockmannbaumgardt08}; \citealt{lockmann09a}; G12)
\begin{equation}
T_{\rm cusp}=\frac{M_{\rm SMBH}}{M_{\rm cusp}}\,{}T_{\rm orb}\,{}f(e),
\end{equation}
where $M_{\rm SMBH}$ is the mass of the SMBH, $T_{\rm orb}$ is the orbital period of a disc star, $M_{\rm cusp}$ is the mass of the cusp inside the stellar orbit, and $f(e)=\frac{1+\sqrt{1-e^2}}{\sqrt{1-e^2}}$ is a function of the eccentricity $e$  of a disc star. 
The main effect of this precession is pericentre advance (e.g. \citealt{subrhaas12}). 

(ii) The disrupted molecular cloud settles into a dense and irregular gas disc. 
The newly formed gas disc and the pre-existing stellar disc exert mutual torques on each other, and thus precess about each other. A star orbiting a SMBH of mass $M_{\rm SMBH}$ 
with semi-major axis $a$,  at an inclination $\beta{}$ relative to a gas disc of radius $R_{\rm DISC}$ and mass $M_{\rm DISC}$ precesses on a timescale (\citealt{nayakshin05}; \citealt{lockmann08}; \citealt{subr09}):
\begin{equation}\label{eq:eq2}
T_{\rm DISC}\sim{}\frac{T_{\rm K}}{\cos{\beta}}\,{}\frac{\sqrt{1-e^2}}{1+\frac{3}{2}e^2}, 
\end{equation}
where
\begin{equation}
T_{\rm K}\equiv{}\frac{M_{\rm SMBH}}{M_{\rm DISC}}\,{}\frac{R_{\rm DISC}^3}{a^{3/2}\,{}\sqrt{G\,{}M_{\rm SMBH}}}.
\end{equation}
$T_{\rm DISC}$ is the timescale for the change of the longitude of the node, i.e. for the precession around the symmetry axis of the gaseous disc (\citealt{subrhaas12}). $T_{\rm DISC}$ depends on the inclination between the two discs, and in particular $T_{\rm DISC}$ tends to infinity if $\beta{}\approx{}90^\circ{}$: the precession will be very slow if the two discs are perpendicular. Instead,  $T_{\rm DISC}$ is minimised, or in other words, the precession occurs at its maximal rate, if the two discs are coplanar. Furthermore, the precession timescale depends on the distance of the star from the SMBH ($T_{\rm DISC}\propto{}a^{-3/2}$). Thus, stars with different $a$ will precess with different speed.

 If $\beta{}=0$ and the two discs are razor thin, then the precession does not affect the inclinations, as it is just a change of the longitude of the node. Instead, if $\beta{}\gtrsim{}5^{\circ{}}$, this precession significantly influences the inclinations of stellar orbits (\citealt{lockmann08}; \citealt{subr09}). 
In the latter case, the orbits of the outer stars will become inclined with respect to the orbits of the inner stars, producing a warp in the disc, and increasing its thickness.  

 Kozai oscillations are another possible effect induced by the interaction between two discs. They trigger a periodic increase of the eccentricity together with small variations of the inclination, and their characteristic timescale is $\propto{}T_{\rm K}\,{}\sin^{-2}{\beta{}}$ (see \citealt{subr09}). Kozai oscillations are efficiently damped in presence of a spherical cusp, if $T_{\rm K}>T_{\rm cusp}$ (which is the case of our run B2, see Fig.~\ref{fig:fig8}).
 As our integrator is not sufficiently accurate for large eccentricities, we select a cloud orbit with $\alpha{}=0$, so that the final inclination between  gas disc and stellar disc remains mild. Thus,  Kozai oscillations are negligible\footnote{As we discussed in the previous Section, the gas disc is not a rigid body, but a collection of concentric annuli with rather different inclination. On the other hand, we exclude that Kozai oscillations play a role, as we do not see any difference in the eccentricity evolution between run~B1 (in which there is no stellar cusp) and run~B2 (where Kozai oscillations should be damped by the stellar cusp).}.

Finally, relativistic effects are not included in our code. Thus, we cannot account for the relativistic precession of the orbits due to the SMBH. The timescale for relativistic precession is (see \citealt{gualandris09}):
\begin{equation}
T_{\rm RP}=\frac{2\,{}\pi{}\,{}c^2\,{}(1-e^2)a^{5/2}}{3\,{}(G\,{}M_{\rm SMBH})^{3/2}},
\end{equation}
where $c$ is the speed of light.

\begin{figure*}
  \center{
    \epsfig{figure=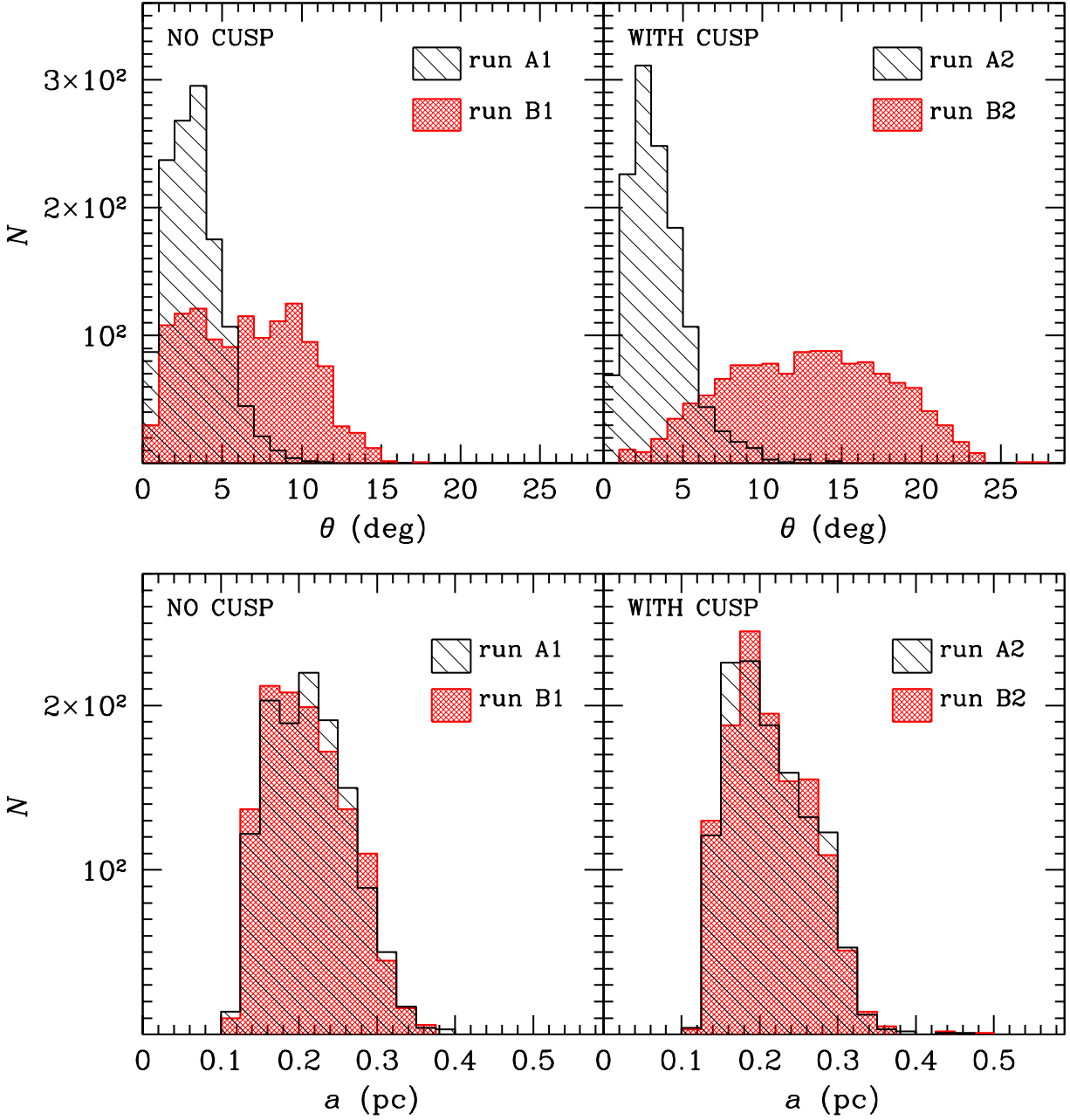,width=17.0cm} 
  \caption{  \label{fig:fig9}
  Top (bottom): distribution of inclinations (semi-major axes) at $t=1.5$ Myr. The inclinations are measured by the angle $\theta{}$, between the normal vector to the orbit of a single star and the total angular momentum vector of the stellar disc (see Appendix~A). Left-hand panel: run~B1 (cross-hatched red histogram) and run~A1 (hatched black histogram). Right-hand panel: run~B2 (cross-hatched red histogram) and run~A2 (hatched black histogram).}}
\end{figure*}

Fig.~\ref{fig:fig8} shows a comparison of the relevant timescales for our simulations. It is apparent that the timescale for relativistic precession is  $\gtrsim{}3$ orders of magnitude longer than the cusp precession timescale. Thus, relativistic precession is negligible for our simulated stellar disc, and the fact that we do not include relativistic effects in our simulations is not an issue.  Fig.~\ref{fig:fig8} shows that $T_{\rm cusp}$ is almost independent of the semi-major axis, for the potential well of the stellar cusp in the GC (see G12). On the contrary, $T_{\rm K}$, which measures the perturbation by the gas disc, depends on the semi-major axis, and is about one order of magnitude shorter for the outermost simulated stars than for the innermost ones.  $T_{\rm K}$ is always longer than $T_{\rm cusp}$. 
 On the other hand, the timescale of disc precession ($T_{\rm DISC}$)
  depends on the angle $\beta{}$ between the stellar orbit and the gas disc.  Furthermore, to derive $T_{\rm K}$ in Fig.~\ref{fig:fig8}, we assumed that the disc is regular, that its mass is $4\times{}10^4$ M$_\odot$, and  that its radius is 0.5 pc. Each of these assumptions is a source of uncertainty, as the disc is not regular, the radius is not clearly defined and the mass depends on the radius. Finally, Fig.~\ref{fig:fig8} is derived in the case of run~B2. The other runs have small differences, which depend on the eccentricity distribution. In the next section, we will consider all the runs in detail.

\subsection{The evolution of the stellar disc in the simulations}\label{starev}
In this section, we investigate the effect of the perturbations generated by the gas disc on the stellar disc, in our simulations B1 and B2. 

In agreement with the fact that the precession cannot affect the semi-major axis of the orbit (from energy conservation, e.g. \citealt{subr09}), the distribution of semi-major axes is not perturbed by the second cloud (bottom row of Fig.~\ref{fig:fig9}). Instead, the distribution of inclinations is significantly affected by the gas disc, in both run~B1 and~B2 (top row of Fig.~\ref{fig:fig9}). In fact, the precession induced by the presence of the gas disc is maximum when $\beta{}$ is small (see equation~\ref{eq:eq2}), and its effect on inclinations is not suppressed by the presence of the stellar cusp (see e.g. \citealt{subr09}). Actually, if the gas disc was perfectly razor-thin and coplanar with the stellar disc ($\beta{}=0$), we would not have found significant changes in the inclinations (e.g. \citealt{subr09}), but just the precession of the ascending node. On the other hand, as we showed in Section~\ref{gasev}, the simulated gas disc is neither razor thin nor perfectly coplanar with the stellar disc. Values of $\beta{}$ as high as $5-20^\circ{}$ occur in some of the streamers, and lead to the increase of the inclinations in a rather complex way.

The fact that stellar inclinations in run~B2 are generally larger than those in run~B1  can be attributed to the different potential well. In fact, in run~B2 the cloud is disrupted marginally faster than in run~B1, as the potential well of run~B2 is the combination of the SMBH and of the stellar cusp. For this reason, the stars in run~B2 are exposed to the influence of the gas disc for a longer time than the stars in run~B1. Also, stochastic fluctuations leading to slightly different inclinations of the streamers might play an important role. 

 Table~2 summarizes some of the main statistical properties of the simulations. 
In run~B1 (B2), the average inclination passes from 2.4$^\circ{}$ (2.4$^\circ{}$) at $t=0$ to 6.6$^\circ{}$ (12.9$^\circ{}$) at $t=1.5$ Myr since the beginning of the simulation (corresponding to $t=1.98$ Myr since the beginning of  run~E in M12). The maximum achieved inclination is 17.3$^\circ{}$ and 27.2$^\circ{}$ in run~B1 and B2, respectively. 

We find no statistically significant differences for the average semi-major axes and for the average eccentricities. We also checked whether or not the  orbital properties  of the simulated stars depend on the stellar mass, and we found no statistically significant difference (at more than 1$\sigma{}$) between stars with mass $>5$ M$_\odot{}$ and stars with mass $<5$ M$_\odot{}$. 

\begin{table}
\begin{center}
\caption{Main statistical properties of the simulations at $t=1.5$ Myr.} \leavevmode
\begin{tabular}[!h]{lllllll}
\hline
Run
& $\langle{}a\rangle{}$
& $\langle{}e\rangle{}$
& $\langle{}\theta{}\rangle{}$
& $a_{\rm max}$
& $e_{\rm max}$
& $\theta{}_{\rm max}$\\

& (pc)
&
& ($^\circ{}$)
& (pc)
& 
& ($^\circ{}$)\\

\hline
\noalign{\vspace{0.1cm}}
A1       & $0.21\pm{}0.05$ & $0.3\pm{}0.1$ & $3.3\pm{}1.7$ & 0.40 & 0.6 & 11.9 \\
A2       & $0.21\pm{}0.05$ & $0.3\pm{}0.1$ & $3.4\pm{}1.9$ & 0.47 & 0.7 & 14.4 \\
\noalign{\vspace{0.1cm}}
B1       & $0.21\pm{}0.05$ & $0.3\pm{}0.1$ & $6.6\pm{}3.5$ & 0.37  & 0.6 & 17.3 \\
B2       & $0.21\pm{}0.05$ & $0.3\pm{}0.1$ & $12.9\pm{}5.0$ & 0.49 & 0.8 & 27.2 \\
\noalign{\vspace{0.1cm}}
G12      & $0.21\pm{}0.05$ & $0.3\pm{}0.2$ & $3.5\pm{}2.1$ & 0.49 & 0.7 & 13.8 \\
\noalign{\vspace{0.1cm}}
\hline
\end{tabular}
\end{center}
\footnotesize{ Average semi-major axis ($\langle{}a\rangle{}$),  average eccentricity ($\langle{}e\rangle{}$), average inclination ($\langle{}\theta{}\rangle{}$), maximum semi-major axis ($a_{\rm max}$), maximum eccentricity ($e_{\rm max}$) and maximum inclination ($\theta{}_{\rm max}$) in our simulations, at $t=1.5$ Myr. The provided uncertainty on the average values is the standard deviation. 
In the last row, we report the simulation presented by G12, for comparison with our run~A2.
The initial values are the same for all the simulations (by construction), and are $\langle{}a\rangle{}=0.21\pm{}0.04$ pc, $\langle{}e\rangle{}=0.29\pm{}0.04$, $\langle{}\theta{}\rangle{}=2.4\pm{}1.5^\circ{}$, $a_{\rm max}=0.39$ pc, $e_{\rm max}=0.44$ and $\theta{}_{\rm max}=7.5^\circ{}$. \\

} 
\end{table}
\begin{figure}
  \center{
    \epsfig{figure=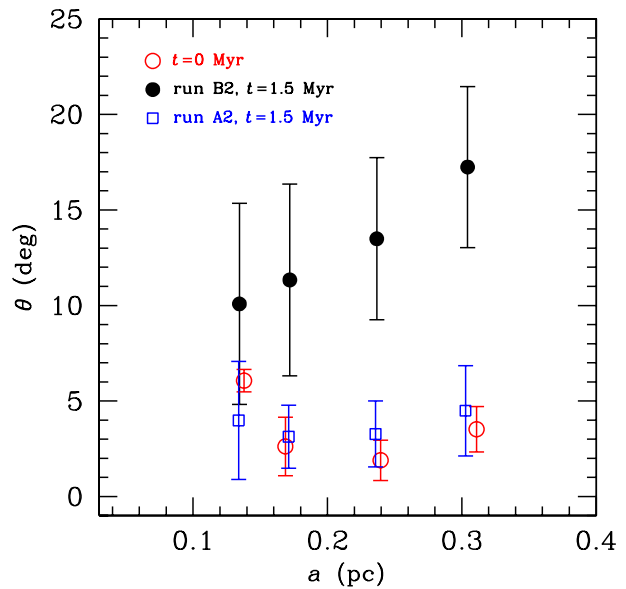,width=8.5cm} 
  }
  \caption{  \label{fig:fig10}
Inclination (with respect to the direction of the average angular momentum) as a function of the semi-major axis in the initial conditions (open red circles), in run~A2 at $t=1.5$ Myr (open blue squares) and in run~B2 at $t=1.5$ Myr (filled black circles). The inclinations of the individual stellar orbits were re-binned in four different bins (each bin was chosen to contain at least 25 stars). Each point is the average inclination per bin, while the error bars show the standard deviation for each bin. 
}
\end{figure}

 In Fig.~\ref{fig:fig10}, the inclinations are shown as a function of the semi-major axis for run~A2 (without gas disc) and run~B2 (with gas disc), at $t=1.5$ Myr. For comparison, we report the initial conditions. In Fig.~\ref{fig:fig10}, the inclinations of the individual stellar orbits were re-binned in four different bins (so that each bin contains at least 25 stars). In the initial conditions, the average inclinations in each bin are well below $7^\circ$. The inner stellar orbits appear more inclined than the outer ones, because the gas disc in run~E of M12 formed with this feature (see M12 for details). In run~B2, the average inclinations ($>10^\circ$) are significantly higher than those in the initial conditions, and it is apparent that the outer stellar orbits achieve (on average) higher inclinations than the inner ones. This is an effect of the precession induced by the gas disc: as $T_{\rm K}\propto{}a^{-3/2}$, the outer orbits are expected to precess faster than the inner ones. As a further support that this is a genuine effect of precession, induced by the gas disc, the inclinations in run~A2 (without gas) remain almost the same as they are in the initial conditions. Furthermore, Fig.~\ref{fig:fig10} indicates that the half-opening angle of the stellar disc at $t=1.5$ Myr is as large as $\sim{}15^\circ{}$ in run~B2, while it does not exceed  $\sim{}7^\circ{}$ in run~A2. Because of this semi-major axis dependent change in the inclinations, the simulated stellar disc in run~B2 (and, similarly, in run~B1) may appear tilted and/or warped.

Recent observations (\citealt{bartko09}, 2010; \citealt{lu09}; \citealt{do13}; \citealt{lu13}) show that the opening angle of the CW disc is only $\sim{}10^\circ{}-14^\circ{}$, but about half of the early-type stars in the inner $1-10$ arcsec ($0.04-0.4$ pc) do not belong to the CW disc. Furthermore, the probability of early-type stars being members of the CW disc decreases with increasing projected distance from Sgr~A$^\ast{}$ (\citealt{bartko09}; \citealt{lu09}). Finally, the CW disc does not seem a flat structure, but rather a significantly warped ($\sim{}64^\circ{}$, \citealt{bartko09}) and tilted object.

Our simulations suggest a reasonable interpretation for such observations: the precession exerted by a slightly misaligned gas disc enhances the inclinations of the outer stellar orbits with respect to the inner stellar orbits. Thus, while the inner disc remains quite coherent, the outer stellar orbits change angular momentum orientation till they may even lose memory of their initial belonging to the same disc. The result is a tilted/warped disc, which is being dismembered in its outer parts. By the time our simulations were stopped ($1.5$ Myr), the warping is still lower than the observed effect. On the other hand, it is reasonable to expect that in few Myr the simulated disc will match the observed features. 

A similar interpretation was already proposed by \v{S}ubr et al. (2009; see also \citealt{haas11a}, 2011b), which pointed out the possible influence of the CNR on the stellar disc. Our paper supports the results  by  \citet{subr09}, by means of the first self-consistent N-body/SPH simulations.
 Other studies considered similar effects induced by a second stellar disc (\citealt{lockmann09a}, 2009b). Alternatively, radiation pressure instability or the Bardeen-Petterson effect were also invoked as possible sources of warping, both driven by the central SMBH (\citealt{ulubay13}). It is likely that the complex distribution of early-type stars in the GC is due to the concurrence of different proposed mechanisms.

\subsection{One or two gas discs?}\label{issue}
 One of the most simplistic assumptions of our simulations is that no gas is left in the initial disc, as we instantaneously removed the remaining gas disc in run~E of M12. This is necessary to prevent our simulations from stalling before the infall of the second cloud. On the other hand, we do not know whether the feedback from the SMBH and/or from the newly born stars is sufficiently efficient to remove all the gas in $\approx{}0.5$ Myr (e.g. \citealt{alexander12}). For a typical young star cluster with mass $\sim{}5\times{}10^3$ M$_{\odot{}}$ and radius $\sim{}0.5$ pc, gas expulsion is expected to occur in $\sim{}10^5$ yr, because of the photo-ionizing flux from O stars (\citealt{kroupa02}). However, the case of the young stellar disc in the GC is more tricky, as the disc is dominated by the gravitational field of the SMBH\footnote{The radiation pressure exerted by a star on a hydrogen atom overcomes the gravitational pull by the SMBH only when the distance of the hydrogen atom from the star is $d_\ast{}\le{}10^{-3}\,{}(L_\ast/10^{39}\textrm{ erg s}^{-1})^{1/2}\,{}(M_{\rm SMBH}/3.5\times{}10^{6}\textrm{ M}_\odot{})^{-1/2}\,{}\,{}d_{\rm BH}$, where $L_\ast$ is the luminosity of the star and $d_{\rm BH}$ is the distance of the hydrogen atom from the SMBH.} and of the stellar cusp. Furthermore, the gas in the disc has a density of $\sim{}10^{7-8}$ cm$^{-3}$, implying that the Str\"omgren sphere around a O star is only $\sim{}10^{-3}$ pc. Dynamical effects (e.g. \citealt{mathews69}; \citealt{franco90}) might increase the radius of the ionized region around early-type stars, but it is more likely that the gas disc can be completely evaporated only by the first supernovae in the stellar disc ($\gtrsim{}3$ Myr). Finally, a possible contribution from the SMBH or other ionizing sources in the GC is even less constrained.

In summary, our assumption of instantaneous removal of the gas from the first disc, although necessary to avoid serious computational-time issues, might underestimate the gas evaporation timescale. This does not alter the dynamical evolution dramatically, as the gravitational influence of the SMBH on the stellar disc is larger by more than two orders of magnitude with respect to that of the gas disc. However, the first gas disc, if not completely removed, is expected to contribute to the secular evolution of the stellar disc. This effect is not included in our simulations.

  On the other hand, our simulations provide important information, even if the actual evaporation timescale is longer. In this case,  our simulations can be considered as a robust lower limit to the effect of a second gas disc. In fact, if the first gas disc was not completely evaporated by the time the second gas disc forms, the effects of the latter may combine with to those of the former, 
 and the resulting precession induced on the stellar disc is expected to be even stronger. 

In the extreme case of almost no gas evaporation in 2 Myr, 
 the results of our simulations mimic the secular influence exerted on the stellar disc by its parent gas disc, rather than the perturbations induced by a second gas disc. This implies that even a long-lived ($>1$ Myr) parent gas disc can significantly affect the inclinations of the stellar orbits.

\section{Conclusions}
\label{sec:concl}
In this paper, we study the effect of perturbations caused by a molecular cloud spiralling towards the GC on the orbits of a pre-existing stellar disc.  It is likely that fresh gas spiralled to the GC in the last few Myrs, as \citet{yusef-zadeh13} observe the traces of very recent star formation ($10^{4-5}$ yr) in the inner parsec.

In our N-body/SPH simulations, the pre-existing stellar disc was formed by a first molecular cloud, disrupted by the GC (run~E of M12). We simulate a second molecular cloud that falls towards the GC, is disrupted by the SMBH, and settles into a dense gas disc. The gas disc perturbs the orbits of the stellar disc. The details of this perturbation depend on the orbital parameters of the two discs and on the presence of a stellar cusp.  The gas disc is neither razor-thin nor regular: it is a collection of different streamers with slightly different inclinations. The inclination of the angular momentum vector of the single streamers with respect to the angular momentum vector of the stellar disc ranges from $\sim{}0$ to $\sim{}20^\circ{}$. Thus, the dynamical interplay between the stellar disc and the gaseous disc(s) is highly complex.


Our simulations show that, if the gas disc and the stellar disc are slightly misaligned ($\sim{}5-20^\circ{}$), the precession induced by the gas disc leads to a significant increase of the inclinations of the stellar orbits. The inclinations of the outer stellar orbits increase more than those of the inner ones, as the precession is faster for larger semi-major axes. This has crucial implications for the evolution of a stellar disc in the GC. 

In fact, the probability of early-type stars being members of the CW disc decreases with increasing projected distance from Sgr~A$^\ast{}$ (\citealt{bartko09}; \citealt{lu09}). Furthermore, the CW disc does not seem a flat structure, but rather a significantly warped ($\sim{}64^\circ{}$, \citealt{bartko09}) and tilted object. 

Our simulations suggest that the outer stellar orbits lose coherence with their original disc because of differential precession. This has two main effects: (i) the stellar disc appears warped/tilted;  (ii) the outer parts of the stellar disc are progressively dismembered. This is consistent with previous predictions by \citet{subr09}, which proposed that the CNR is the source of differential precession.

In our simulations, we assumed that the parent gas disc instantaneously evaporated, and that the naked stellar disc interacts with a second molecular cloud. It is likely that the stellar disc is still embedded in some residual of the original gas disc. Thus, our simulations provide a robust lower limit to the effect of a second gas disc. In fact, if the first gas disc was not completely evaporated, the new disc `superimposes' to the old one, and the resulting precession induced on the stellar disc is expected to be even stronger.





Another effect expected to be important is Kozai resonance. In our simulations, Kozai resonance is negligible as the timescale for Kozai oscillations scales as $\propto{}T_{\rm K}\,{}\sin^{-2}{\beta{}}$.  Kozai resonance becomes important only if $\beta>>0$, but might be completely suppressed by the stellar cusp (e.g. \citealt{subr09}).  The case with $\beta\sim{}90^\circ{}$ deserves to be investigated with a more suitable integration scheme. Finally, we do not observe fragmentation in the simulations presented in this paper, but the mass resolution of gas particles is a factor of 10 lower than in M12. 
Thus, the possibility that the second gas disc  fragments into self-bound clumps, forming stars, will be addressed  in a further study, with higher resolution simulations. 



\section*{Acknowledgments}
We thank the referee, Ladislav \v{S}ubr, for his valuable comments that significantly improved our paper. We also thank H. Perets, E. Ripamonti, L. Mayer, R. Alexander, W. Dehnen, A. King  and S. Nayakshin for useful discussions. We thank the authors of \textsc{gasoline} (especially J. Wadsley, T. Quinn and J. Stadel). We acknowledge the CINECA Award N. HP10CL51UF, 2012 for the availability of high performance computing resources and support.  MM acknowledges financial support from the Italian Ministry of Education, University and Research (MIUR) through grant FIRB 2012 (`New perspectives on the violent Universe: unveiling the physics of compact objects with joint observations of gravitational waves and electromagnetic radiation'), and  from INAF through grant PRIN-2011-1 (`Challenging Ultraluminous X-ray sources: chasing their black holes and formation pathways').

\begin{appendix}
\section{The coordinate system}
\begin{figure}
  \center{
    \epsfig{figure=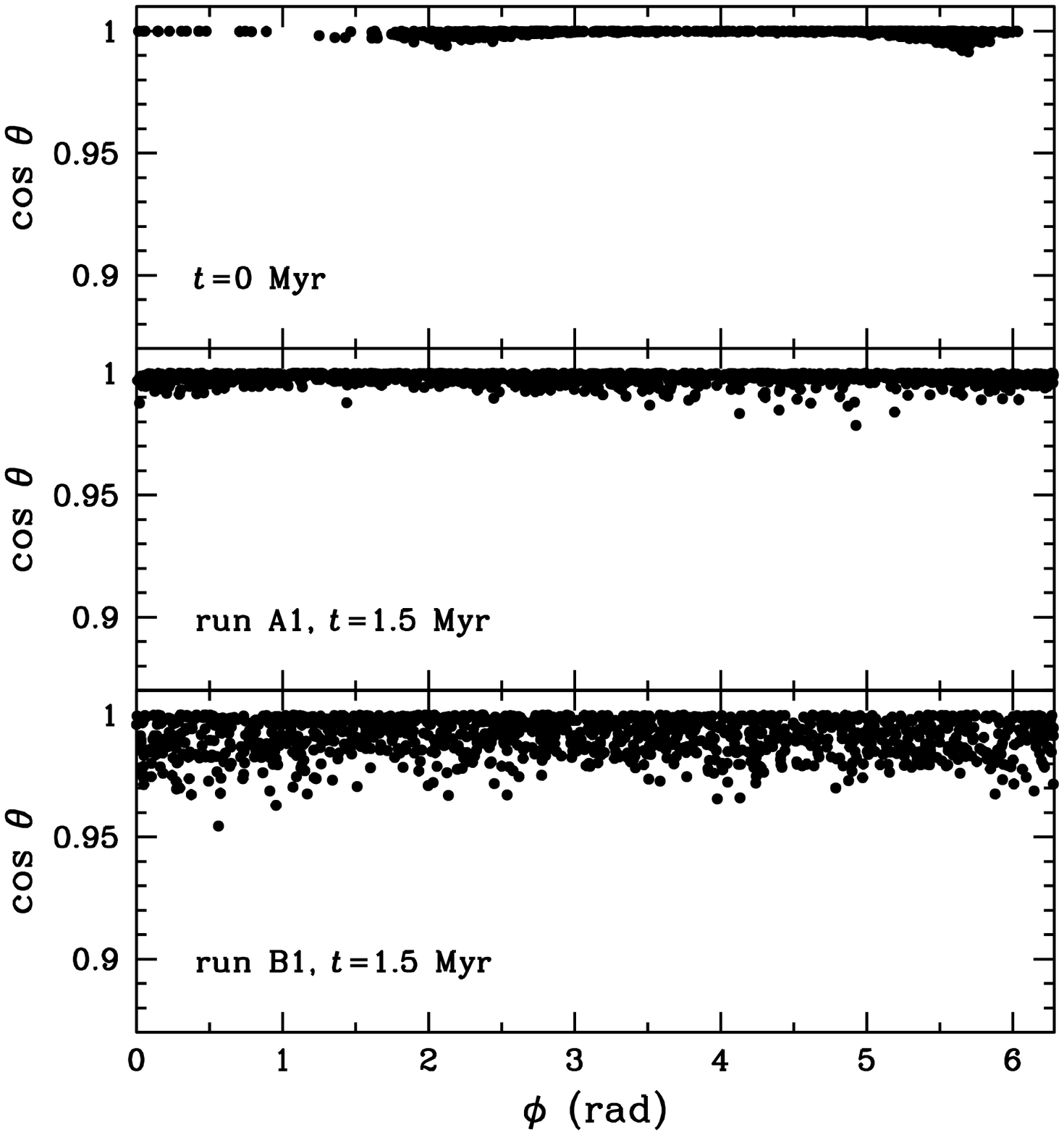,width=8.5cm} 
  }
  \caption{  \label{fig:figA1}
Each point marks the position of a stellar orbit in the plane defined by  $\cos{\theta}$ and $\phi{}$ (see equation~\ref{eq:A2}). Top panel: initial conditions (the same for all runs). Central panel: run~A1 at $t=1.5$ Myr. Bottom panel: run~B1 at $t=1.5$ Myr. 
}
\end{figure}

\begin{figure}
  \center{
    \epsfig{figure=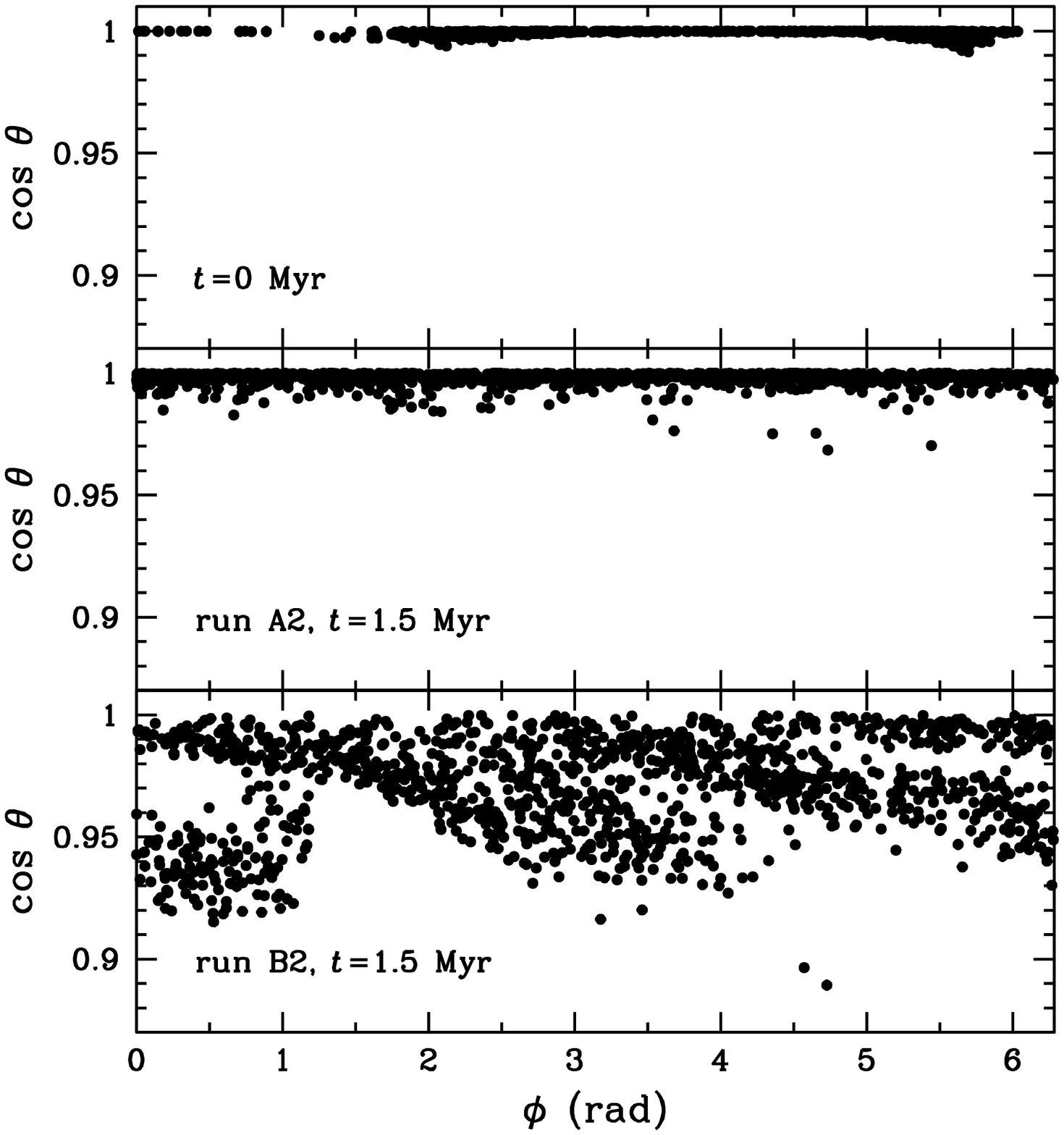,width=8.5cm} 
  }
  \caption{  \label{fig:figA2}
Each point marks the position of a stellar orbit in the plane defined by  $\cos{\theta}$ and $\phi{}$ (see equation~\ref{eq:A2}). Top panel: initial conditions (the same for all runs). Central panel: run~A2 at $t=1.5$ Myr. Bottom panel: run~B2 at $t=1.5$ Myr. 
}
\end{figure}

The coordinate system is defined in the following way. We define ${\bf J}_{\rm DISC}$ as the total angular momentum of the stellar disc. The normal vector to the total angular momentum of the stellar disc will then be 
\begin{eqnarray}
{\bf n}_{\rm DISC}={\bf J}_{\rm DISC}/J_{\rm DISC}=\nonumber{}\\
=(\cos{\phi{}_{\rm DISC}}\,{}\sin{\theta{}_{\rm DISC}},\,{}\sin{\phi{}_{\rm DISC}}\,{}\sin{\theta{}_{\rm DISC}},\,{}\cos{\theta{}_{\rm DISC}}),
\end{eqnarray}
where $J_{\rm DISC}$ is the modulus of ${\bf J}_{\rm DISC}$. We now define the coordinate system so that $\cos{\theta{}_{\rm DISC}}=1$ at any given time $t$. The normal vector to the angular momentum of a single star will then be  
\begin{equation}\label{eq:A2}
{\bf n}={\bf J}/J=(\cos{\phi{}}\,{}\sin{\theta{}},\,{}\sin{\phi{}}\,{}\sin{\theta{}},\,{}\cos{\theta{}}), 
\end{equation}
where ${\bf J}$ and $J$ are the angular momentum vector of a single star orbit and its modulus, respectively. By construction, $\theta{}=\theta{}_{\rm DISC}=0$ means that the angular momentum of a star is aligned to the total angular momentum of the disc. Thus, we refer to the value of  $\theta{}$ as the inclination of the stellar orbit with respect to the disc axis at a given time $t$. The average value of $\theta{}$ ($\langle{}\theta{}\rangle{}$) gives also a measure of the half-opening angle of the disc. Fig.~\ref{fig:figA1} shows how the stellar orbits populate the plane defined by $\cos{\theta}$ and $\phi{}$ in the initial conditions and in runs~A1 and B1. Fig.~\ref{fig:figA2} is the same as Fig.~\ref{fig:figA1}, but for runs~A2 and B2.
\end{appendix}
\end{document}